\documentclass[pre,preprint,showpacs]{revtex4}

\usepackage{amsmath}

\usepackage{graphicx}

\usepackage{float}
\usepackage[english]{babel}
\usepackage{cleveref}
\usepackage{natbib}

\begin{document}

\title{The Tension on dsDNA Bound to ssDNA-RecA Filaments May Play an Important Role in Driving Efficient and Accurate Homology Recognition and Strand Exchange}

\author{Julea Vlassakis}
\author{Efraim Feinstein}
\author{Darren Yang}
\author{Antoine Tilloy}
\author{Dominic Weiller}
\author{Julian Kates-Harbeck}
\author{Vincent Coljee}
\author{Mara Prentiss}
\affiliation{Harvard University, Department of Physics, Cambridge, MA, 02138}
\email{prentiss@fas.harvard.edu}

\begin{abstract}

It is well known that during homology recognition and strand exchange the double stranded DNA (dsDNA) in DNA/RecA filaments is highly extended, but the functional role of the extension has been unclear. We present an analytical model that calculates the distribution of tension in the extended dsDNA during strand exchange. The model suggests that the binding of additional dsDNA base pairs to the DNA/RecA filament alters the tension in dsDNA that was already bound to the filament, resulting in a non-linear increase in the mechanical energy as a function of the number of bound base pairs.  This collective mechanical response may promote homology stringency and underlie unexplained experimental results.

\end{abstract}

\maketitle

\section{Introduction}
Sexual reproduction and DNA damage repair often include homologous recombination facilitated by RecA family proteins \cite{Cox,SteveK}. In homologous recombination, a single stranded DNA molecule (ssDNA) locates and pairs with a sequence matched double-stranded DNA molecule (dsDNA). In the first step of the process, the incoming ssDNA binds to site I in RecA monomers, resulting in a helical ssDNA-RecA filament with 3 base pairs/monomer and $\sim 6$ monomers/helical turn \cite{Pavletich}. This helical ssDNA-RecA filament then searches dsDNA molecules for homologous sequences by rapidly binding and unbinding dsDNA to site II in RecA \cite{Pavletich}. Thus, the sequence of the ssDNA in the searching filament is fixed, and the system then searches through the available dsDNA to find a sequence match for that ssDNA. The binding of dsDNA to site II is very unstable, so if the dsDNA is not homologous to the ssDNA bound to site I, the dsDNA rapidly unbinds from the ssDNA-RecA filament. If the dsDNA is homologous, 
strand exchange should occur, probably via base-flipping that transfers the Watson-Crick pairing of the complementary strand from the outgoing DNA strand bound to site II to the incoming DNA strand bound to site I \cite{Radding}. This strand exchange reduces the unbinding rate for the dsDNA \cite{Singleton}. RecA is an ATPase, but \textit{in vitro} homology recognition and strand exchange can occur without ATP hydrolysis \cite{Mazin,Menetski,Stasiak}. Thus, each step in the homology search/strand exchange process is fully reversible.

\begin{figure}[H]
\centering
\includegraphics[width=88 mm]{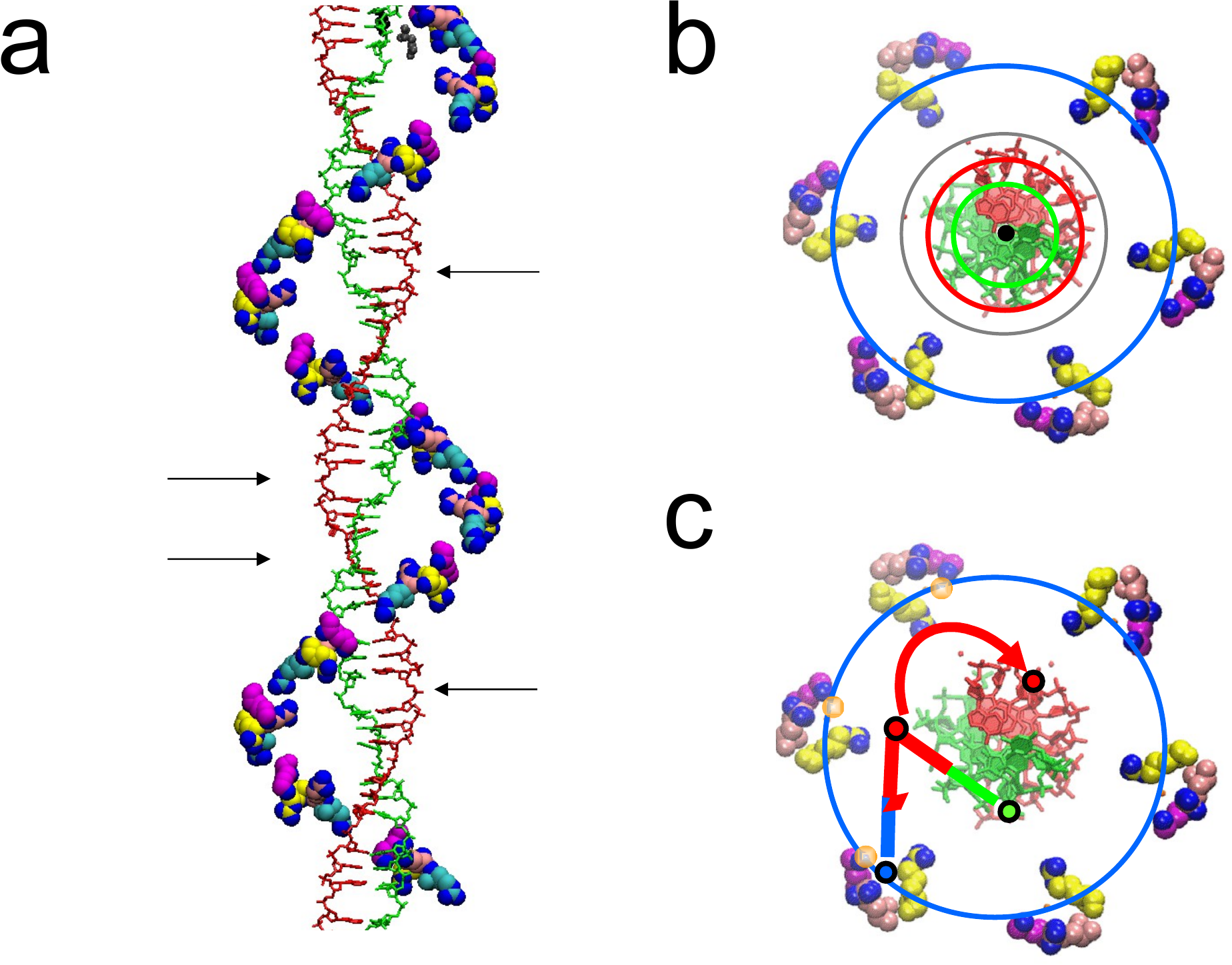}
\caption{(Color Online) a. Side view of the X-ray structure of the dsDNA post-strand exchange (final) state with complementary (on the right at the arrow) and incoming strands  shown as green and red stick renderings. The arrow points to a  few rises between two triplets in the dsDNA structure. The VMD (32) renderings of RecA crystal structure 3CMX (8) show site II residues Arg226 (pink), Lys227 (cyan), Arg243 (yellow), and Lys245 (magenta) with charged nitrogen atoms (blue). The cyan triangle indicates the approximate position of an outgoing strand phosphate} b. top view of the same structure with circles indicating the radii occupied by the incoming (green), outgoing (blue), and complementary strands (red for final state and gray for intermediate state). c. top view showing the base pairing in the homology recognition/strand exchange process superimposed on the actual X-ray structure.
\label{fig:crystal}
\end{figure}

\begin{figure}[H]
\centering
\includegraphics[width=88 mm]{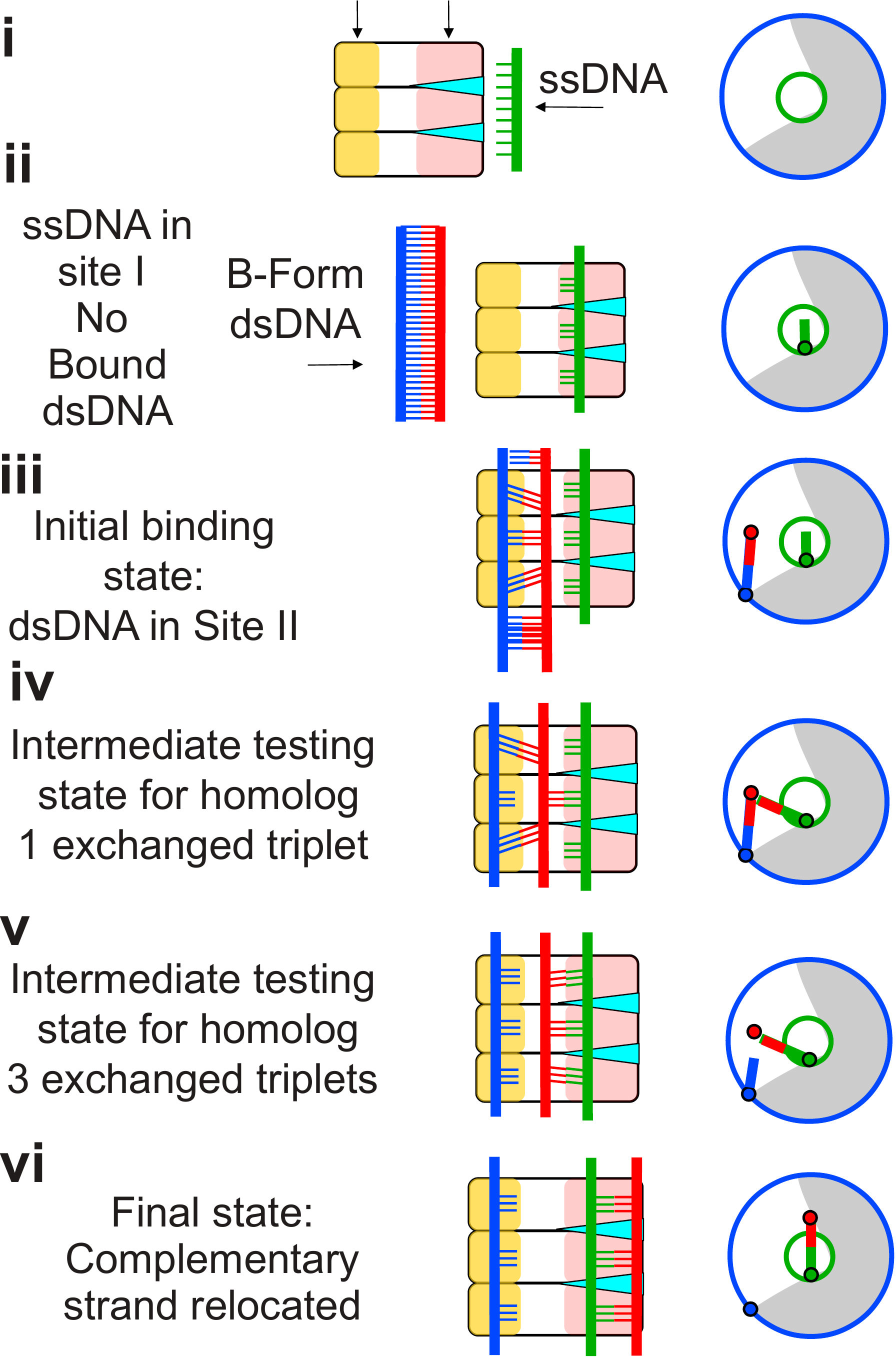}
\caption{(Color Online) Schematic of Interactions between dsDNA and the ssDNA-RecA filament as strand exchange progresses for a homolog with side views shown in the central part of each panel and top views shown in the right part of each panel where the grey region indicates the space occupied by the protein DNA/protein and the circles indicate the radii occupied by the three strands in their final post-strand exchange positions.  In the schematic, the filament consisting of three RecA proteins, showing site I (pale red), site II (pale orange), and the support for the rises provided by the L1 and L2 loops (cyan), with (i) unbound ssDNA ( (ii) ssDNA bound in site I and unbound B-form dsDNA, (iii) ssDNA bound in site I and dsDNA bound in site II; The outgoing strand (far left ), complementary strand (bound to the outgoing strand in purple ), and incoming strand (bound to the protein on the right) are shown in blue, red, and green, respectively. (iv) central triplet has undergone strand exchange, resulting in a 
decrease in lattice mismatch and a decrease in the stress on the bp (v) all three triplets shown have undergone strand exchange (vi) ssDNA bound in site II and dsDNA bound in site I in the final assembled state which has even less bp stress than the strand exchanged state due to increase mechanical support for the rises.}
\label{fig:RecAbindingpicmp}
\end{figure}

During the homology search and strand exchange process, dsDNA bound to RecA is extended significantly beyond the B-form length \cite{Koller}. Recent theoretical work proposed that the free energy penalty associated with extension may promote rapid unbinding of non-homologous sequences, but the free energy penalty was assumed to be a linear function of the number of bound triplets and the kinetic trapping due to near homologs was not considered \cite{Tlusty}.Earlier work had also suggested that the dsDNA extension promotes base-flipping \cite{MazinSteveK} and reduces kinetic trapping since the lattice mismatch between extended dsDNA and B-form dsDNA presents a steric barrier to interactions between unbound dsDNA and bound ssDNA which implies that the dsDNA must bind to the filament in order to interact with the ssDNA \cite{Bruinsma}. These studies assumed the dsDNA in the DNA/RecA filament is uniformly extended; however, the X-ray crystal structures of the dsDNA in the final post-strand exchange state and the 
ssDNA in the homology searching state both consist of base pair triplets in a nearly B-form conformation separated by large rises as illustrated in Figure~\ref{fig:crystal}a. The rises occur at the interfaces between adjoining RecA monomers  \cite{Pavletich}, as illustrated in   Fig.~\ref{fig:RecAbindingpicmp}. The functional role of the non-uniform extension has been unclear.

In this paper we present a simple model that calculates the extension of each base pair triplet in a dsDNA.  Using this model, we calculate the free energy changes associated with progression through the homology recognition/strand exchange process. The results of that calculation suggest a resolution to the long standing question of why strand exchange is free energetically favorable even though the Watson-Crick pairing in the initial and final states is the same and the DNA/protein contacts in the ssDNA-RecA filament and final post-strand exchange state are nearly the same.\cite{Pavletich} The model also makes several significant qualitative predictions, the most significant being the suggestion that  the collective behavior of the triplets due to their attachment to the phosphate backbones leads to a free energy that is a non-linear function of the number of consecutive bound triplets. As a result of this non-linearity, total binding energy has a minimum as a function of the number bound triplets in a 
given conformation. After that minimum is reached, adding more triplets that  given conformation becomes free energetically unfavorable.

Such a change in sign in the binding energy as a function of the number of bound triplets can never occur in a theory where the energy is a purely linear function of the number of bound triplets since the binding of any base pair anywhere in the system is equally likely regardless of the state of any of the other triplets in the system.  In a system with a binding energy that is a linear function of the number of correctly paired bound triplets, if homologous triplets can initially bind to the system, then additional homologous triplets will always continue to bind.  Thus, binding will readily progress across a non-homologous triplet.  As we will discuss in detail in this work, in a system with a linear energy and more than $\sim 4$ binding sites, either homologs will be too unstable or near homologs will be too stable.

In contrast, the non-linearity  may provide more rapid and accurate homology recognition than is available for systems using linear energies because the non-linearity requires that  dsDNA binding to the ssDNA-RecA  proceed iteratively triplet by triplet through a series of checkpoints which inhibit the progression of strand exchange past a non-homologous triplet.    At each checkpoint, the progression of strand exchange to more stable binding conformations is only free energetically favorable if a sufficient number of contiguous homologous base pair triplets are bound to the ssDNA/RecA filament in the appropriate conformations.  The general qualitative features of the homology recognition based on the non-linear energy  follow from basic properties of the simple model and are insensitive to the parameters chosen.  These features include the following: 1. that homology recognition will proceed iteratively through consecutive triplets 2. that strand exchange reversal is much more favorable at the ends of 
the filament than at the center 3. that there will be two checkpoints that cannot be passed unless the bound dsDNA contains a sufficient number of contiguous homologous base pairs in the appropriate conformations.   Though the general features of the model are very robust, the exact number of contiguous homologous bp required to progress past a particular checkpoint depends strongly on the choice of model parameters.  Analytical modeling and numerical simulations suggest that there are a small range of parameters that allow the free energies predicted by the model presented in this paper to provide homology recognition which is both fast and accurate   \cite{Julian, Efraim}.

Simulation results suggest that though the initial binding of $\sim 9$ base pairs (bp) is free energetically accessible,  adding more bound triplets is not favorable, and adding more than 15 bp is enormously unlikely unless the first checkpoint is passed.  \cite{Efraim} The first major checkpoint requires that $\sim 9$ of the $\sim 15$ bp that initially bind to the filament are contiguous and homologous.   If the initial  $\sim 15$ bp do not contain   $\sim 9$ contiguous homologous base pairs, the dsDNA cannot make a transition to the more stably bound intermediate state; therefore, the weakly bound dsDNA will almost immediately unbind from the filament.  This checkpoint rapidly rejects all but 50 of the $\sim$ 10,000,000 possible binding positions in a bacterial genome.  If the initial  $\sim 15$ bp do include $\sim 9$ contiguous homologous base pairs, the system can make a transition to a metastable intermediate state, which allows more  base pairs to be added to the filament.   The second major checkpoint 
occurs when $\sim 18$ contiguous bp are bound to the filament in the metastable intermediate state.  If all of the bp are contiguous and homologous, the system can make a transition to the final post-strand exchange state.  Otherwise, the long regions of accidental homology  will slowly reverse strand exchange and unbind.  Given the statistics of bacterial genomes, passing the homology requirement for the  second checkpoint would guarantee that the correct match had been found.  These predictions are in good agreement with known experimental results that measure the stability of strand exchange products as a function of the number of contiguous bound base pairs \cite{Hsieh}.

In this work, we will not attempt to optimize the parameters of the model in order to provide rapid and accurate  homology recognition.    Rather, we will consider why homology recognition systems in which  the energies  are a linear function of the number of bound base pairs can either provide rapid unbinding of non-homologs or stable binding of complete homologs, but not both if the number of binding sites in the system is $> \sim 4$ .  We will then discuss how qualitative features of the non-linear free energy predicted by the model allow strand exchange to avoid kinetic trapping in near mismatches while also permitting homologs to progress completely to strand exchange in systems where the number of binding sites is $>4$.

\subsection{General Issues in Self-Assembly Based on the Pairing of Arrays of Matching Binding Sites}

In efficient self-assembly/recognition systems that create correct assembly by matching linear arrays of binding sites, correctly paired arrays of binding sites must remain stably bound, whereas incorrectly paired arrays must rapidly unbind even if the incorrect pairing contains only one single mismatched binding site.  For a system in thermodynamic equilibrium, the populations in different binding configurations are determined simply by their binding energies.  Thus,  in a system where every array consists of $N$ binding sites if $U(m,N)$ is the binding energy when $m$ of the binding sites are correctly paired, accurate recognition requires that  $\exp[-(U(m,N)-U(N,N))/(kT)] \ll 1 ~~~\forall~m<N$, where $k$ is Boltzmann's constant and $T$ is the Kelvin temperature.  Furthermore, in a system with a temperature $T$, the requirement that mismatched pairings rapidly unbind implies that $U(m,N)>-kT ~~~\forall~m<N$, whereas if the correctly bound ones are to remain stably bound $U(N,N)$ must be $\ll -kT$. If the 
energy is a linear function of the number bound and a mismatch contributes zero free energy  then $U(N,N)= ~-~N \epsilon kT$  and $U(N-1,N)= ~-~ (N-1) \epsilon kT$, so the condition $\exp[-(U(N-1,N)-U(N,N))/(kT)] \ll 1 $ requires $\epsilon \gg 1$, which also implies that the homolog will remain stably bound.  For simplicity consider $\epsilon =3$ , which implies $U(N,N)= - N \epsilon kT = - 3 N kT$ . Substitution into the requirement for the unbinding of near homolog requires that and $U(N-1,N)= - (3N-3)  kT >-kT ~\implies N<4/3  $ Thus, the requirements for rapid and accurate recognition can only be met if $N=1$. The requirements become more stringent if the specificity ratio is more strict than 1/20.   As we will discuss below, accurate homology recognition in a bacterial genome requires accurate recognition over a length of more than 12 bp, which implies $N>4$ since 12 base pairs is 4 triplets.

Some of the problems with realizing accurate recognition in systems at thermodynamic equilibrium were recognized by John Hopfield in the 1970's, inducing him to propose a kinetic proofreading system that requires an irreversible process. \cite{Hopfield}  In such systems, the energy of the bound state can be very deep without making the energy of the searching state deep because of the irreversible step that transfers the system from the searching state to the bound state.  Even in Hopfield's system there is a tradeoff between search speed and accuracy since greater sequence discrimination requires greater unbinding probabilities for homologs.  The increased binding probability for homologs increases the searching time because the correct binding site must be revisited many times before the homolog makes the irreversible transition to the bound state.  Earlier work had proposed that RecA based homology recognition could proceed via kinetic proofreading,~\cite{Singleton},  but homology recognition \textit{in vitro}  
is known to proceed without an irreversible step. \cite{Mazin,Menetski,Stasiak}   In this work, we will consider the non-linearities in the free energy as a function of the number of contiguous bound base pairs that arise as a function of the differential extension of dsDNA bound to RecA and how those non-linearities can promote rapid and accurate homology recognition without requiring irreversibility by making transitions between successive bound states contingent on the state and relative position of the bound base pairs.

\subsection{Homology Recognition and Strand Exchange are Likely to Proceed in Units of Base pair Triplets}

Previous experimental results suggest that homology recognition occurs via base flipping of the complementary strand bases between their initial pairing with the outgoing strand and their final pairing with incoming strand. \cite{Singleton,Bazemore,Alex}.  Experimental results have already shown that strand exchange does progress in triplets.  \cite{Ragunathan} We propose that within each triplet the nearly B-form structure preserves sufficient stacking to allow homology discrimination to exploit the energy difference between the Watson-Crick pairing of homologs and heterologs, while the large rises between triplets result in mechanical stress that plays several important functional roles in homology recognition and strand exchange.  Furthermore, if the stacking makes it free energetically favorable for the triplets to flip as a group, the lost of Watson-Crick pairing due to a single mismatch in a triplet may make strand exchange unfavorable for the entire triplet, as would be the case if Watson-Crick 
pairing alone determined the free energy of the strand exchanged state.  If a single mismatch makes strand exchange of a triplet unfavorable, then  testing in triplets implies that in a random sequence the possibility of an accidental mismatch is 1/64, whereas if the search were done by comparing single base pairs the probability of an accidental match would be 1/4.  Homology stringency and searching speed  would be greatly reduced if the search were not done in triplets.  Once a sufficient number of base pair triplets have undergone strand exchange, the complementary strand backbone relocates to the position shown in the X-ray structure of the final state \cite{Singleton, Alex}, as illustrated in Figures~\ref{fig:crystal}c and~\ref{fig:RecAbindingpicmp}.

\subsection{General Model for dsDNA Bound to a RecA Filament}

Certain underlying assumptions and qualitative features are important to the model proposed in this paper: 1. that in the absence of hydrolysis the extension of the protein filament is unaffected by dsDNA binding \cite{Karplus} 2. that homology recognition and strand exchange occur in quantized units of  base pair triplets \cite{Pavletich,Ragunathan} 3. that the incoming and outgoing strands  consist of nearly B-form triplets separated by large rises where the strand is bound to the filament due to strong charged interactions between the backbone phosphates and positive residues in the protein \cite{Pavletich,Danilowicz,Alex} 4. that the complementary strand is bound to the filament dominantly via Watson-Crick pairing, resulting in large stress on the base pairs unless the dsDNA is in the final post-strand exchanges state where the L1 and L2 loops provide significant mechanical support. \cite{Pavletich,Dekker}

The model utilizes work presented by deGennes that calculated the force required to shear dsDNA \cite{deGennes}. We extend deGennes' model  to the triplet structures in the initial, intermediate, and final dsDNA conformations.   In the model the actual three dimensional helical structure is converted into a one dimensional system. In the simple one dimensional model, $L_{R}$  is rise between the triplets in the incoming and outgoing strands and   a single variable, $\gamma$, characterizes the equilibrium spacing between phosphates when the complementary strand is in a particular state.  Thus, for a particular state, the difference between the equilibrium spacing  is given by   $(1-\gamma )L_R$ .

\subsection{Predicted Extension and Energy }

\begin{figure}[H]
\centering
\includegraphics[width=88 mm]{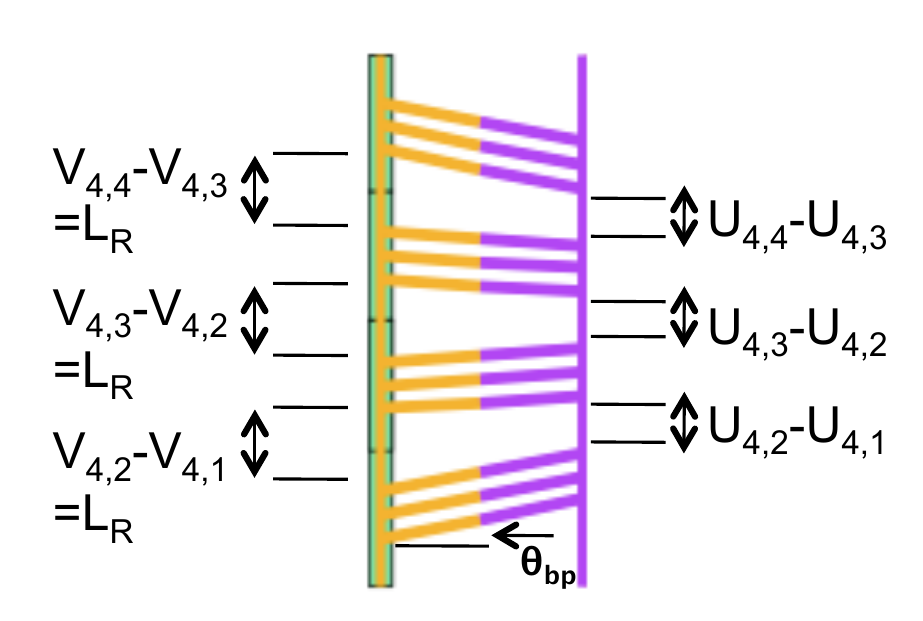}
\caption{Schematic of dsDNA bound to site II of the protein showing the model parameters.}
\label{fig:model}
\end{figure}

As shown in Fig.~\ref{fig:model}, the extensions of rises between base pair triplets in a strand bound directly to the protein  are given by $v_{N,i}$ and for a system with $N$ triplets bound RecA,
\begin{equation}
\label{eq:v}
v_{N,i} = i L_{R}.
\end{equation}

The $u_{N,i}$  specify the extensions of the rises in the complementary strand.  At equilibrium,  the net force on  each $u_{N,i}$ must be zero; therefore,  for $j = 2$ to $N$
\begin{equation}
\label{eq:zeroforce}
Q((u_{N,j+1} - u{N,j}- \gamma L_{R}) - (u_{N,j} - u_{N,j-1} - \gamma L_{R})) + R(u_{N,j} -v_{N,j}) = 0,
\end{equation}
where $R$ and $Q$ are the spring constants for the base pairs and the backbones, respectively. These values for $R$ and $Q$ may be substantially different from those for individual dsDNA base pairs when the dsDNA is not bound to RecA because of the  interactions between the charged phosphates and the protein and because  dsDNA is grouped in triplets where the stacking between the triplets is strongly disrupted; however, it is still likely that $R \ll Q$ as it is in naked dsDNA since the interactions between the bases on opposite strand  is significantly weaker than the interaction between the phosophates in the backbone of the same strand.

The boundary condition on the last triplet $u_{N,1}$ requires that
\begin{equation}
\label{eq:boundary}
+ Q(u_{N,2} - u_{N,1} - \gamma L_{R}) + R(u_{N,1} - v_{N,1}) = 0
\end{equation}

The values of $u_{N,i}$ can be found using Equations~\ref{eq:v}-\ref{eq:boundary}. The angle between base pairs and the DNA helical axis is shown as $\theta_{bp}$ in Fig.~\ref{fig:model}.

In the continuous limit where the discrete subscript $i$ is replaced by a continuous variable $x$, the equations have an analytical solution
\begin{equation}
\label{eq:tilloy1}
v_{N,x}= x +A_N(x) \sinh( \chi x/L_R)
\end{equation}
where $\chi=Sqrt[R/(2Q)]$ is the deGennes length for RecA bound dsDNA and  the constant $A$ is found by using the boundary conditions for the ends, yielding
\begin{equation}
\label{eq:tilloy2}
A_N(x) \Big[R \sinh(\chi x/L_R) + Q \chi \cosh(\chi x/L_R)\Big] = Q(\gamma -1  ) L_R.
\end{equation}

In the limit where $1/(2 \chi) \gg 1$
\begin{equation}
\label{eq:tilloy3}
A_N(x)  = { (\gamma -1  ) L_R\over{   \cosh(\chi N/2)}}.
\end{equation}

\begin{figure}[H]
\centering
\includegraphics[width=88 mm]{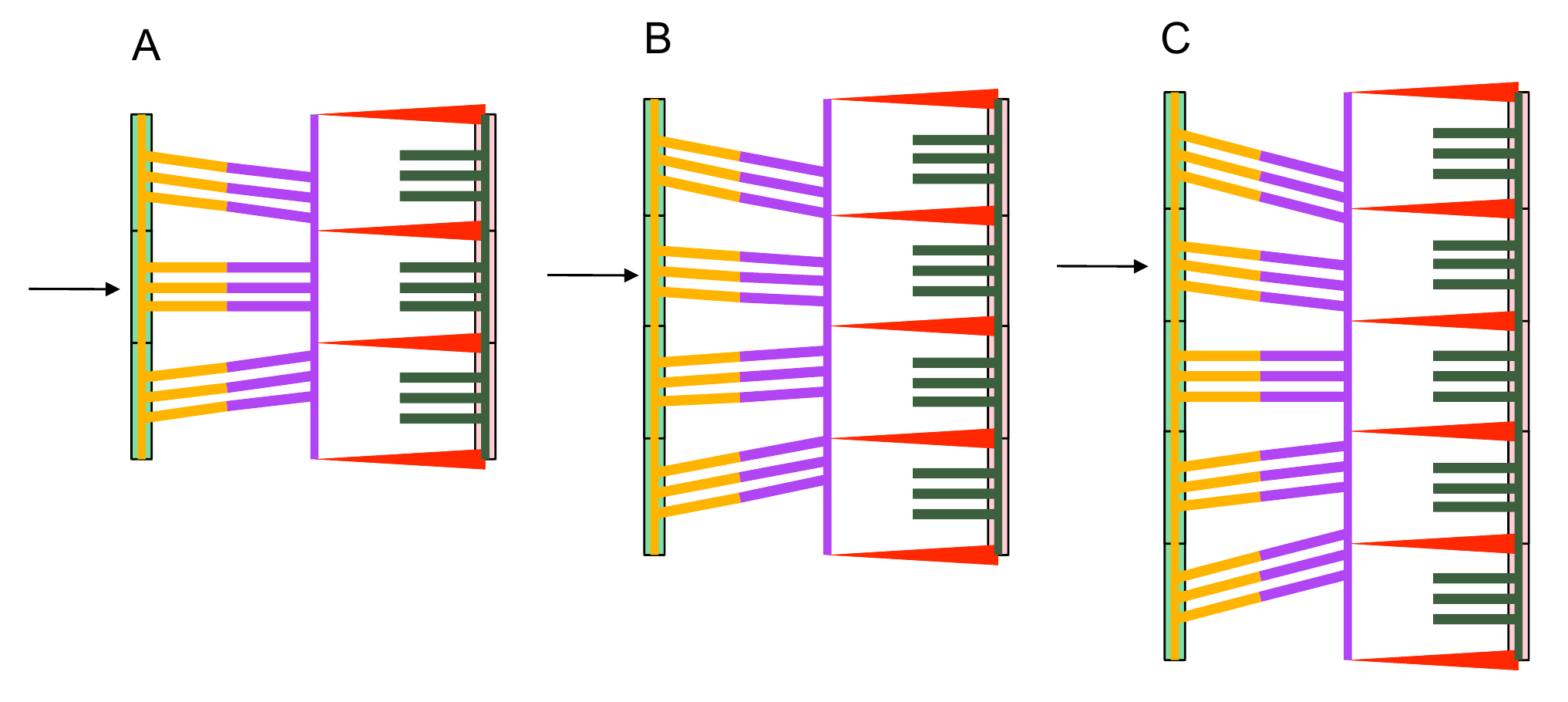}
\caption{Effects of adding triplets to the initial bound state as calculated by the model.  (A)  Three triplets shown in the bound state with an arrow pointing to the second one.  (B)  Adding a triplet to the initial state changes the tension in the other triplets.  (C)  Again tension on the triplets change as a fifth triplet is added to the initial state.}
\label{fig:tripletaddition}
\end{figure}

\begin{figure}[H]
\centering
\includegraphics[width=88 mm]{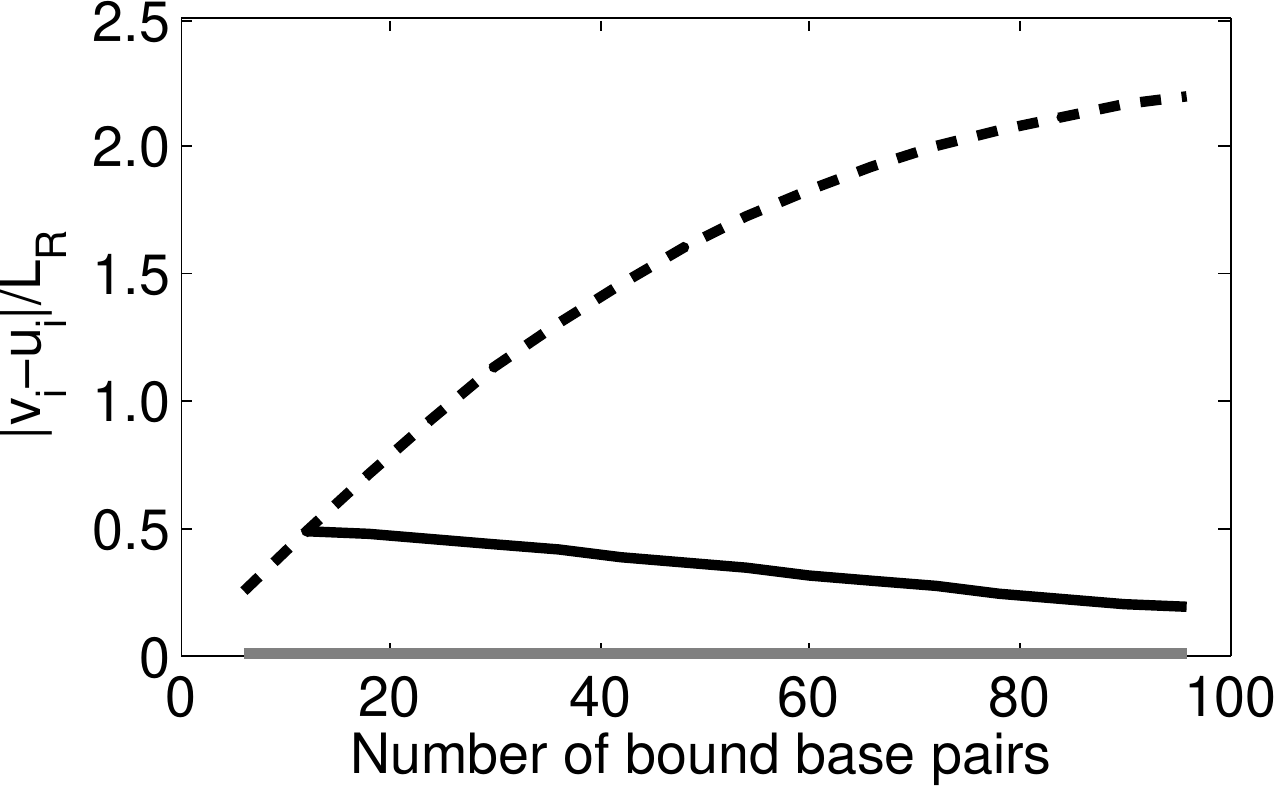}
\caption{The dashed, solid gray, and solid black lines show the absolute value of $v_{i} - u_{i}$ for the outer triplet, central triplet, and the first triplet out from the center respectively when the number of bound triplets is odd.}
\label{fig:rises}
\end{figure}

These assumptions and features lead to a nuanced picture of the distribution of tension during strand exchange. The lattice mismatch between the complementary strand and its pairing partners is largest at the ends of the filament as shown in Fig.~\ref{fig:tripletaddition}; therefore, the base pair tension is largest at the end of the filament.  Furthermore,  the lattice mismatch at the ends increases significantly with the number of bound triplets as shown in Fig.~\ref{fig:tripletaddition} and \ref{fig:rises}. The angle $\theta_{bp}$ is greatest at the ends of the molecule.    When fewer than $\sim 30 $ bp are bound to the filament, the tension on the base pairs at the ends  increases rapidly as more RecA bound triplets are added, as shown in Fig.~\ref{fig:tripletaddition} and \ref{fig:rises}; however,  $\theta_{bp}$  and the tension on the base pairs near the center decreases as more triplets are added, as shown in Fig.~\ref{fig:tripletaddition} and~\ref{fig:rises}. In contrast with the tension on the base 
pairs, which is largest at the ends of the complementary strand, the tension on the rises in the complementary strand is largest at the center and smallest at the ends. These general qualitative features are not sensitive to $R$, $Q$ or $\gamma$ as long as $R \ll Q$.

Given  the values of $u_{N,i}$ the mechanical energy of the system, which is of the form $1/2 k(x-x_{0})^2$, may be calculated from:
\begin{equation}
\label{eq:Emech}
E_{mech}(N) = [\sum\limits_{i=1}^N \frac{1}{2} R(u_{N,i}-v_{N,i})^2 +\sum\limits_{i=2}^N \frac{1}{2} Q(u_{N,i}-u_{N,i-1} - \gamma L_{R})^2]
\end{equation}

When $R \ll Q$ this expression simplifies to:
\begin{equation}
\label{eq:Emech2}
E_{mech}(2) = 2 [\frac{1}{2}R \frac{(L_{R}(1-\gamma)}{2}^2]
\end{equation}

\begin{equation}
\label{eq:Emech3}
E_{mech}(3) = 2 [\frac{1}{2} R(L_{R}(1-\gamma)^2]
\end{equation}

These energy terms represent the stress on base pairs at the ends of the molecule due to the lattice mismatch. When $R \ll Q$,   for moderate numbers of bound base pairs, this energy is stored primarily in the extended base pairs.  This idea is particularly important because it implies that an increase in the extension of the rises in the complementary strand can reduce the free energy by reducing the tension on the base pairs even if the increase in extension of the rises requires some energy. Similarly, in the RecA structure the transition from the intermediate state to the final state may reduce the tension on the base pairs because the interactions between the base pairs and  the L1 and L2 loops  may  increase the equilibrium extension of the rises in the complementary strand.

Using the continuous limit allows us to generate scaling laws for the extension as a function of the total number of bound base pairs. In the continuous limit,  the non-linear contribution to the free energy is given by
\begin{equation}
\label{eq:analytical energy}
 E_{non-linear}(N)= \frac{1}{2} \frac{ (1- \gamma)^2 L_R^2}{\chi^2 \cosh^2[\chi N/2]} \sum_{i=-N/2}^{i=N/2}  \sinh^2[\chi i].
\end{equation}

Thus, when  $\chi \ll 1$ the non linear energy term has the following scaling
\begin{equation}
\label{eq:analytical energy1}
 E_{non-linear}[N] \alpha  (1- \gamma)^2 L_R^2 \frac{ \sinh(\chi N) - \chi N} {1+ \cosh^2[\chi N/2]}.
\end{equation}

In the limit where the number of base pairs bound is much less than the deGennes length so  $\chi N \ll 1$ the energy
the non-linear energy scaling is
\begin{equation}
\label{eq:analytical energy limit }
 E_{non-linear}[N] \alpha  (1- \gamma)^2 L_R^2 \chi^2 N^2.
\end{equation}

Thus, when $N$ is small the energy increases as the square of the number of bound triplets, consistent with the exact results for the discrete case  given in equations \ref{eq:Emech},\ref{eq:Emech2} and~\ref{eq:Emech3}.  In contrast, when $N$ is larger than the deGennes length, the non-linear energy term approaches zero and the mechanical energy increases linearly with increasing $N$.  In this limit, the base pairs at the center of the filament are no longer under tension.  Thus, adding a triplet to the end of the filament effectively adds another triplet to the unstressed center rather than increasing the stress on all of the bound triplets.

The total energy of the system includes the mechanical energy calculated above and the non-mechanical binding energy per RecA monomer, $E_{bind}$.  Assuming the free energy of an unbound dsDNA is zero and the free energy gain upon binding a triplet is independent of $N$ and $i$ then the non-mechanical contribution to the binding energy for $N$ triplets is $N E_{bind}$. When the first RecA monomer binds, $E_{total}[1] = E_{bind}$, which is a constant negative value.  In contrast, when $N>1$, the stress on the molecule yields a total energy of
\begin{equation}
\label{eq:Etot}
E_{total}[N] = E_{mech}[N] + N E_{bind}
\end{equation}
which changes in sign and magnitude depending on $L_{R}$, $R$, $Q$ and $\gamma$.

\subsection{Modeling the Homology Search and Strand Exchange Process}

Recent experimental results suggest that the homology recognition/strand exchange process uses four major dsDNA conformations: 1. B-form dsDNA, which is the structure of the dsDNA when it is not bound to the protein 2. an initial sequence independent searching state with the dsDNA bound to the RecA where the complementary strand bases paired with outgoing strand 3. an intermediate sequence dependent strand exchanged state in which the complementary strand bases are paired with the incoming strand where the complementary strand backbone is in a position where its bases can flip between pairing with the incoming and complementary strands 4. the final state known from the X-ray structure where the complementary strand bases are paired with the incoming strand bases and the phosphate backbone is in the position shown in the x-Ray structure. \cite{Alex}

Consistent with strand exchange proceeding through base flipping of triplets,  recent experimental results have suggested that outgoing strand bases are arranged in B-form triplets separated by large rises such that the complementary strand bases can readily rotate between pairing with the outgoing strand and pairing with the incoming strand. \cite{Danilowicz}.   The crystal structure shows that the incoming strand is located near the center of the helical DNA/protein structure whereas the residues associated with the binding of the outgoing strand are much  farther away from the center,   as illustrated in Fig.~\ref{fig:crystal}; however,  if strand exchange occurs via the base flipping of base pair triplets, the spacing within triplets must be approximately the same for all three strands.  Thus, given that the total extension of the outgoing strand backbone is  be much larger than the total extension of the incoming strand backbone, the rises in the outgoing strand must be  much larger than the rises 
in the incoming strand, as illustrated in Fig.~\ref{fig:RecAbindingpicmp}.

In the one dimensional model considered here where $\gamma$ is the only parameter characterizing each dsDNA conformation, $\gamma$ is smallest for the initial searching state, where the complementary strand is paired with the very highly extended outgoing strand, consistent with experimental results that the dsDNA in the initial bound state has a large differential extension between the outgoing and complementary strands that prevents more base pairs from binding to the filament unless the dsDNA undergoes strand exchange. \cite{Danilowicz}  The $\gamma$ for the intermediate state is slightly larger  because the complementary strand is paired with the less extended incoming strand. This value is  consistent with experimental results that suggest that the strand exchange of homologous triplets is favorable because it reduces the differential tension on the dsDNA. \cite{Alex}   The $\gamma$ for the final state is almost 1, because of the support provided for the rises by the L1 and L2 loops of the protein that 
occupy the rises in the complementary strand, consistent with the known experimental result that the binding of at least $\sim 80$ bp in the final state is free energetically favorable. \cite{Dekker} In order to probe the role of tension during homology recognition strand exchange we apply the model to the various stages of strand exchange using the following parameters:  $\gamma = {0.4,0.75, 0.8, 0.9875}$ and $E_{bind}$ equal to $0, -0.75 kT$, $-0.8 kT$ and $-0.125 kT$ per homologous triplet in the unbound, initial bound, intermediate  and final states, respectively. These $\gamma$ values are inspired by X-ray structure DNA RecA filaments, the known properties of B-form dsDNA, experimental results on the stability of strand exchange products \cite{Hsieh}, and results of numerical simulations that optimize homology recognition. \cite{Efraim}

\section{Results and Discussion}

\subsection{Changes in Mechanical Tension Allows Strand Exchange to be Free Energetically Favorable if and only if 6 Contiguous Homologous bp Undergo Strand Exchange Insensitive to Model Parameters}

Early RecA recognition systems assumed that strand exchange is always free energetically favorable, just as Watson-Crick pairing of unpaired ssDNA is always favorable, but this assumption is not correct. In self-assembly based on  ssDNA/ssDNA pairing,  the pairing of matched bases is free energetically favorable and the pairing of mismatched bases is approximately neutral. Thus, in ssDNA/ssDNA pairing systems, such as DNA origami,  assembly of  matching sites is always free energetically favorable because the correct Watson-Crick pairing reduces the free energy below that for the system where no bases were bound.  In contrast, if only Watson-Crick pairing is considered, self-assembly based on strand exchange is free energetically neutral for matched bases and free energetically unfavorable for mismatched bases because the system begins in a state with all of the bases in the complementary strand correctly paired with the corresponding bases in the outgoing strand. Thus, previous models of strand exchange that 
considered only the Watson-Crick pairing were faced with a paradox: at thermal equilibrium strand exchange is only free energetically neutral, so at best only $50 \%$ of correct pairings would end up in the strand exchanged state.  The other $50  \%$ would end up paired with their initial partners; however, \textit{in vivo} strand exchange proceeds to completion.  Including protein contacts did not solve the problem since the protein contacts in the initial ssDNA-RecA filament are almost identical to the contacts in the final post-strand exchange state \cite{Pavletich}; however, the model presented in this paper resolves the problem, as discussed below.

In the model considered here, the coupling along the dsDNA backbone makes any change in the dsDNA conformation  (such as adding or base-flipping a triplet) alter the positions and extensions of all other bound triplets. This effect is illustrated in Fig.~\ref{fig:tripletaddition}, which shows the rises and extensions of the base pairs calculated from the model when an additional triplet binds in site II of the protein when all of the bound triplets are bound in site II. The triplet highlighted with the arrow experiences increasing tension as triplets are added to the end of the sequence. Furthermore, as shown in Fig.~\ref{fig:rises},  the lattice mismatch at the ends of the filament increases monotonically as more triplets are added.  In contrast, the lattice mismatch  for the first triplet out from the center of the sequence decreases as base pairs are added (solid black line) because the added triplets extend the inner rises.    As the number of bound triplets approaches the deGennes length for RecA bound 
dsDNA, the lattice mismatch at the ends of the filament approaches a constant and the central rises in the complementary strand approach $L_R$.

The mechanical stress model considered here suggests strand exchange is free energetically favorable if and only if strand exchange transfers a rise because the reduction in free energy is the result of the reduction in mechanical stress due to a reduction in the stress on the base pairs due to decrease in the lattice mismatch between the complementary strand and its Watson-Crick pairing partner.  Thus, when the dsDNA first binds to the searching filament and all of the triplets are in the initial state, base flipping of a single triplet does not lower the energy of the system since a rise is not transferred even if that triplet is perfectly homologous; however, once one homologous triplet is strand exchanged, it becomes favorable to transfer a contiguous homologous triplet.  Thus, the transfer of the Watson-Crick pairing of the complementary strand from the outgoing strand to the incoming strand  is not free energetically favorable except in sequence regions containing at least six base pairs of contiguous 
homology.  This minimum number of contiguous homologous bases required to pass the first checkpoint is a basic qualitative feature of the model that is extremely insensitive to the model parameters chosen.  This first checkpoint can provide very rapid unbinding of almost all initial dsDNA pairings, which is highly advantageous for rapid searching.  Though the model specifies that 6 contiguous bases is the minimum number required to pass the checkpoint insensitive to the parameters used in the model,the actual number of base pairs required to pass the first checkpoint  is sensitive to parameters.

Experimental results suggest that strand exchange is only marginally stable when $\sim 9$  contiguous homologous bp undergo strand exchange,\cite{Hsieh} suggesting that rapid unbinding will occur for all tested sequences except for $1/4^{9} \sim 2 \times 10^{-6}$ which represents only $\sim$ 50 possible positions for a bacterial genome with a length of 10,000,000 bp.  All other sequences will rapidly unbind from the RecA filament because the  binding energy for sequence independent searching state is very weak when only $\sim 9$ bp are bound  and adding more base pairs to the searching state would increase rather than decrease the binding energy.

\subsection{Calculations of the Free Energy as Function of the Number of Bound Triplets in Each dsDNA Conformation}

\begin{figure}[H]
\centering
\includegraphics[width=88 mm]{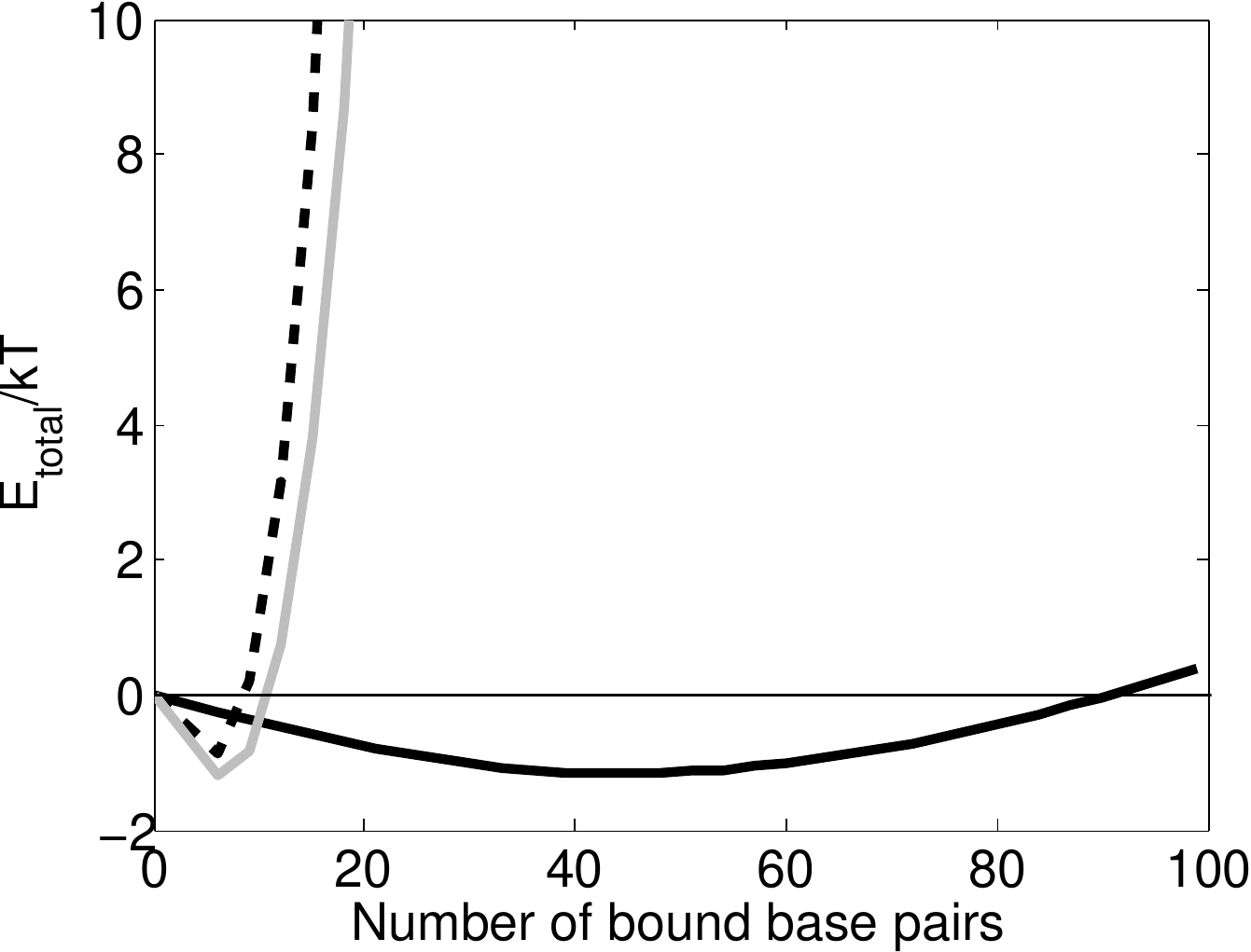}
\caption{Total energy as a function of bound base pairs all in the initial bound state (dashed line), the intermediate state (solid gray line) and the final post-strand exchange state (solid black line).}
\label{fig:nonlinearity}
\end{figure}

In order to understand the progression through homology recognition and strand exchange, it is important to consider the free energies for all of the dsDNA conformations involved, not just the initial bound state and the intermediate state.  Of course the process is kinetic, so the energies of the transition states play a vital role; however, for simplicity  we will only show the energies of the various bound conformations using $\gamma$ values based on numerical simulations. \cite{Efraim} Fig.~\ref{fig:nonlinearity} shows $E_{total}$ as a function of number of base pairs in the initial bound state, the intermediate state and the final post-strand exchange state when all of the triplets are in the same conformation. Except for the initial binding of $\sim 9$ to 15 bp,  all of the triplets are rarely in the same conformation. Thus,  the free energy curves for states with all of the triplets same conformation  rarely represent the free energy of the system; however, it is clear that reductions in the 
mechanical energy of the bound dsDNA can drive strand exchange for homologs, as we discuss below.

\subsection{Non-linearities in the Free Energy Allows Homologs to Progress to Complete Strand exchange}

While it is energetically favorable to add base pairs to the initial bound state for small numbers of base pairs, the quadratic term of $E_{total}$ from Equation~\ref{eq:Emech} rapidly increases as a function of increasing number of base pairs, making the binding of a large number of base pairs in the initial state unfavorable. This is because of the significant tension due to the lattice mismatch between the complementary strand and the outgoing strand; consequently,  for the parameters based on simulation, no more than $\sim 15$  bp can bind to the filament in the initial searching state because the energy required greatly exceeds $kT$. Thus, once $\sim$ 15 bp are bound to the filament, the system is in a highly free energetically unfavorable state which will force it to choose between the following: 1. unbinding from the  filament 2. strand exchange, which is unfavorable for non-homologs.  Again, the prediction that there will be an checkpoint in the progression of strand exchange from the initial bound 
state to the intermediate state is insensitive to the model parameters.   The parameters only determine whether more than 6 contiguous homologous base pairs are required to progress past the checkpoint.

\begin{figure}[H]
\centering
\includegraphics[width=88 mm]{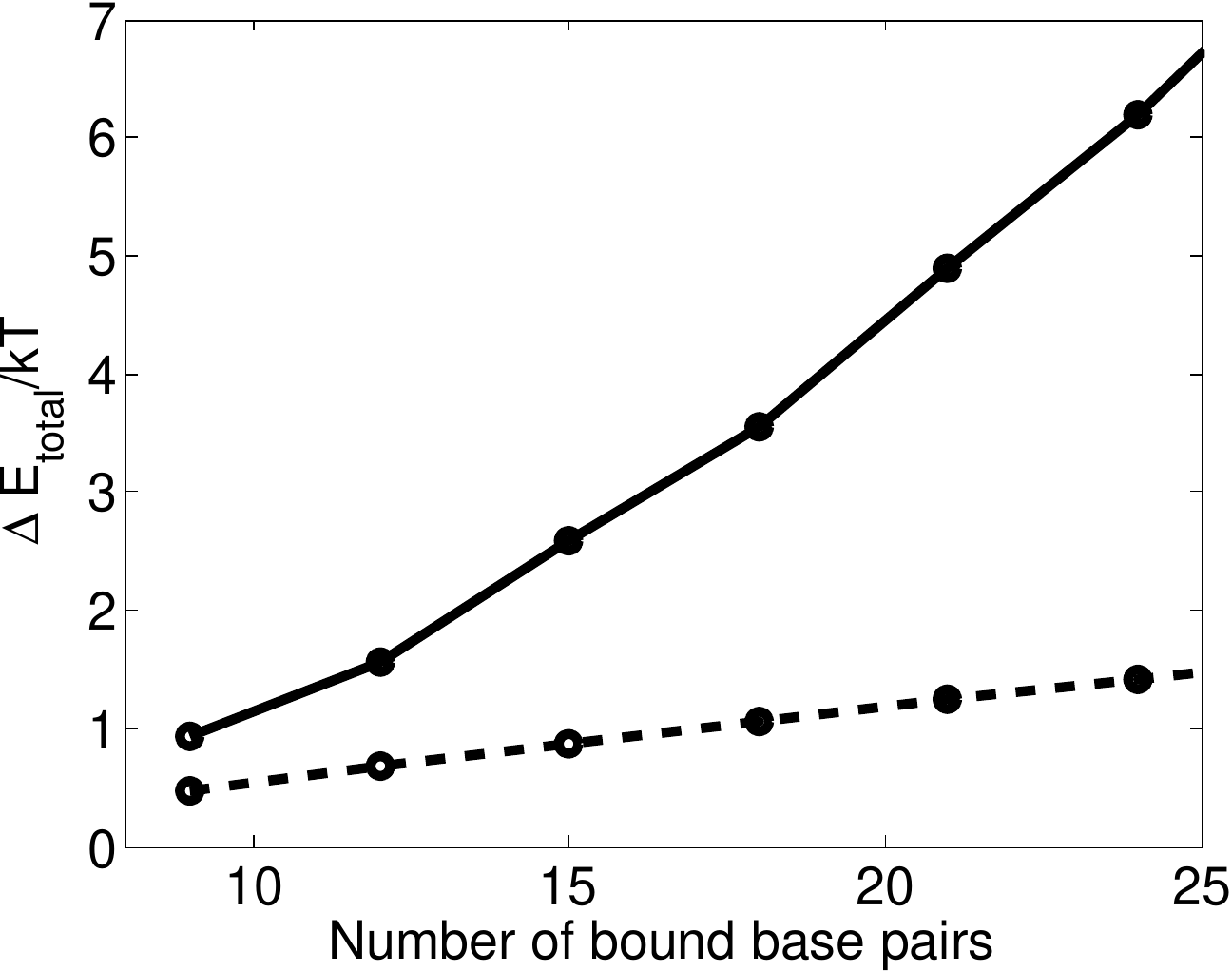}
\caption{For a completely homologous dsDNA with all triplets in the intermediate strand exchanged state, the free energy penalty for a single triplet reversing strand exchange by making a transition from the intermediate strand exchanged state to the initial bound state as a function of then number of base pairs when the base flipping triplet is in the  center of the filament (solid line) or at the end of the filament (dashed line)  }
\label{fig:flippenalty}
\end{figure}

Homologs can rapidly progress to complete strand exchange if the weak initial  binding holds  long enough for $\sim 9$ homologous contiguous base pairs to undergo strand exchange, which stabilizes the binding for those homologous bp and allows strand exchange to progress. \cite{Julian,Efraim}  The non-linearity in the free energy makes the strand exchange of consecutive homologous triplets increasingly favorable as long as the number of bound base pairs is $< \sim$ 30; consequently,   the non-linearity in the free energy makes strand exchange reversal more improbable as a more contiguous homologous base pairs are strand exchanged.  Furthermore, the non-linearity makes strand exchange at the center of the filament  increasingly unfavorable as the number of bound triplets increases as shown in Fig.~\ref{fig:flippenalty}, while still allowing strand exchange reversal to remain possible at the ends of the filament.   Again, these qualitative features are basic properties of the model that are highly insensitive to 
the model parameters.    These qualitative features allow true homologs to progress to complete strand exchange even though non-homologs readily unbind.    In contrast, for a system with a linear free the probability that strand exchange will be reversed for a given triplet is independent of the number of other triplets bound and of the position of the particular triplet in the filament.  As a result, such systems either suffer rapid unbinding of homologs or strong kinetic trapping in near homologs, as discussed above for the general case of a system with a linear binding energy and greater than 4 binding sites.

\subsection{dsDNA Tension Drives the Transition from the Intermediate State to the Final State}

The free energy penalty due to binding  in the intermediate state is so large that even a perfect homolog cannot bind more than $\sim 18$ bp in the intermediate state.  The only way to continue to add base pairs to the filament is to make a transition to the final post-strand exchange state where the L1 and L2 loops provide significant mechanical support for the rises.  This transition reduces the tension sufficiently to allow more triplets to bind in the initial bounds state.  If the transition does not occur, strand exchange will reverse because the intermediate state energy is unfavorable.  Thus,  the collectivity in the behavior of the dsDNA can have  a significant effect in enforcing homology stringency at this final step in the strand exchange process by enforcing the following:   1. that the number of base pairs that can be bound in the intermediate state is limited 2. that transfer of a single triplet from the intermediate state to the final state is never favorable  3. that transferring non-
contiguous triplets from the intermediate state to the final state is unfavorable even if they are homologs 4. that the transfer of pairs of triplets is not favorable until $\sim 18$ bp are bound in the intermediate state.  Properties 1-3 are shared with the transition from the initial bound state to the intermediate state, but the fourth property is different.  It may arise from some combination of two features: 1. the linear energy in the final state may be less favorable than the linear energy in the intermediate state 2. there is a significant boundary penalty associated with the deformation of the backbone that occurs when the system is partially in the intermediate state and partially in the final state.  In either case, then the transition from the intermediate state to the final state will only become free energetically favorable when the favorable non-linear term  becomes dominant over the other terms.  If,  the final state only becomes free energetically favorable when  $\sim 15$ contiguous base 
pairs are present in the final state, then the non-linearity in the energies of the intermediate and final states in the energy may provide additional discrimination against regions of accidental homology since for a particular given searching ssDNA sequence the odds of this occurring with a given searching sequence are $\sim 1.4x10^{-11}$/base pair. Thus, even in a 10 million bp genome, the probability that such an accidental homology is present in a given bacterial genome  is $\sim 1/10^{4}$.   We have considered a few bacterial genomes and found that the sequences are indeed random in this sense, with the exception of repeated genes.  Thus, many bacterial genomes contain no accidental mismatches consisting of 18 contiguous bp; therefore, no further homology checking is required if 18 contiguous bp exactly match.  The important statistical property is consistent with experimental results that strand exchange products to do not become stable until more than $\sim 18$ bp have undergone strand exchange.

\subsection{For dsDNA in the final state the free energy has a minimum as a function of $N$}

If the final state has a much higher $\gamma$ than the intermediate state, then adding base pairs will remain free energetically favorable for much larger $N$; however, as long as the mechanical contribution remains a non-linear function of $N$, the free energy as a function of $N$ will achieve a minimum for some $N$, after which adding more base pairs becomes free energetically unfavorable.  Eventually, as $N$ approaches the deGennes length, the mechanical energy will become a linear function of $N$.  In this case, the energy cost of adding an additional triplet remains constant as a function of $N$.  If this cost is small in comparison with $kT$ and unbinding is forbidden, the model suggests that the length of the strand exchange product can increase without limit.

\subsection{The Effect of the Non-Linearity on the Strand Exchange of a Mismatched Triplet}

It has previously been assumed that the free energy penalty for strand exchange of a triplet is approximately equal to the loss of Watson-Crick pairing for that triplet, with a possible additional factor due to the effect of the mismatch on the pairing of the two neighboring bases which ranges from $\sim 1. 5$ to $\sim  4 kT$ . \cite{SantaLucia}   In contrast, for a system with the non-linearity considered here if the initially bound base pairs contain a single mismatch, then strand exchange may be significantly more unfavorable because the  unfavorable free energy contribution due to this mismatch must include not only the Watson-Crick pairing energy for that base pair and its neighbors, but also  the  the increased mechanical stress on the two matched base pairs. This stress not only makes a direct contribution to the free energy penalty, but it can also increase the stacking penalty by distorting the bonds between the two homologous base pairs which lowers their Watson-Crick pairing energy.  A  detailed 
structural calculation would be required to correctly assess all of these factors.  In what follows, we will assume that the free energy penalty for the strand exchange of a mismatched base is approximately equal to the  Watson-Crick pairing loss as long as the number of $< 18$ bp are bound to the filament.

\subsection{dsDNA Tension Inhibits Progression of Strand Exchange Past a Mismatch}

\begin{figure}[H]
\centering
\includegraphics[width=88 mm]{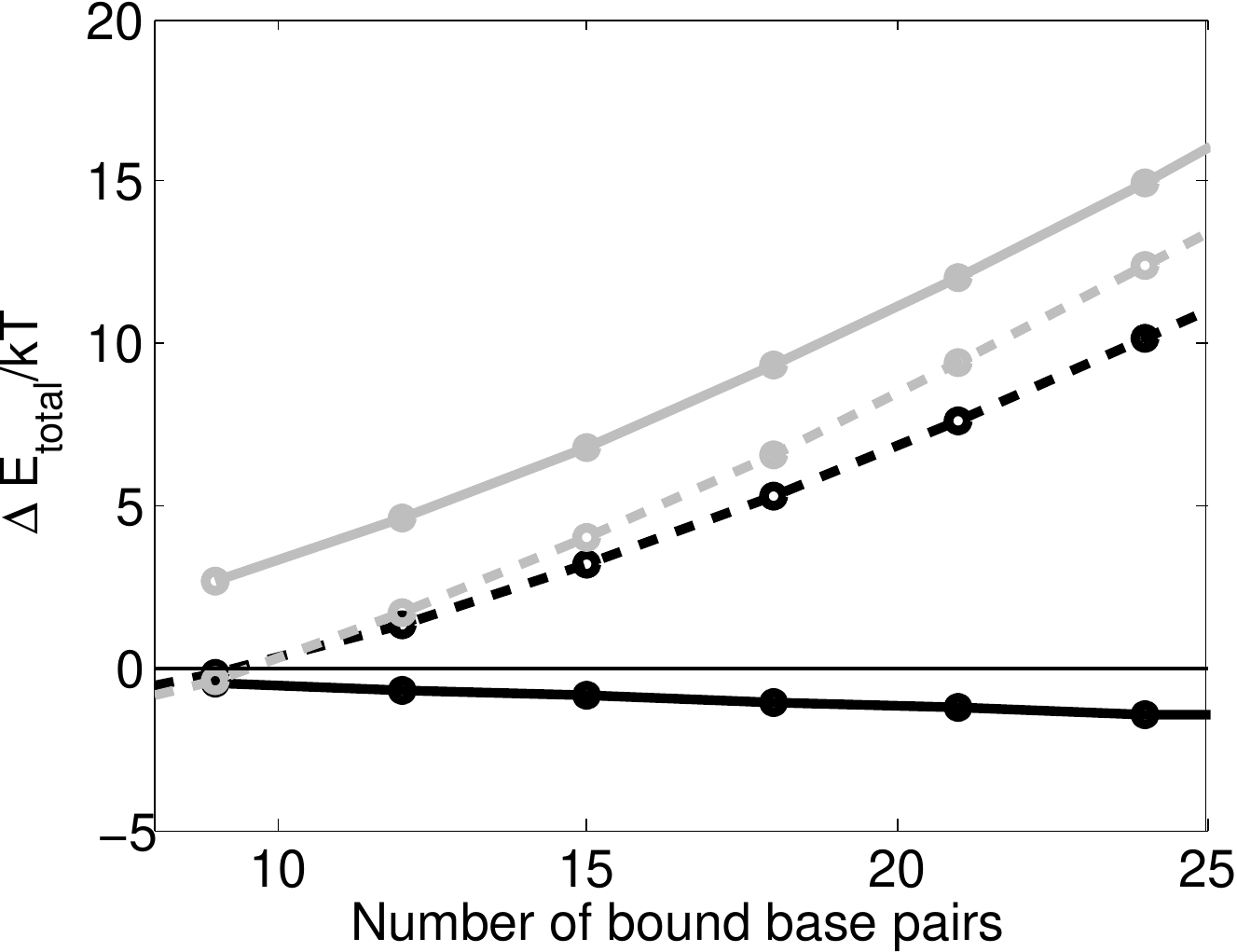}
\caption{The free energy penalty for adding another triplet in the initial bound state as a function of the number of base pairs in the strand exchanged state.  The dashed black line shows the penalty when all of the triplets are homologous.  The dashed gray line shows the penalty for adding an additional triplet to the initial bound state after a triplet that is non-homologous.  The solid gray line shows the penalty for adding a second triplet after a non-homolog where the first triplet after the non-homolog was homologous. The solid black line shows the energetic favorability of strand exchanging a homologous triplet in the initial bound state.}
\label{fig:nonhomologpenalty}
\end{figure}

In a system with a linear free energy as a function of the number of bound homologous triplets, adding more homologous triplets is always favorable even if the last triplet added were non-homologous, resulting in enormous kinetic trapping.  In contrast, the non-linear energy inhibits binding of additional base pairs after a non-homologous triplet has bound, as illustrated in Fig.~\ref{fig:nonhomologpenalty}.   The dashed black line shows the curve for a perfect homolog adding a triplet to the initial bound state if all of the other bound triplets have undergone strand exchange.   For up to $\sim 18$ bp thermal energy is sufficient to bind additional base pairs.  The dashed gray line shows the free energy penalty for adding a homologous triplet if the last triplet added was non-homologous.  The free energy penalty is only slightly larger than the penalty for a homolog; however, the solid gray line shows that the penalty for adding a second triplet is very large, even though both triplets added after the non-homolog were in fact 
homologous. For comparison, the solid black line shows the energetic favorability of strand exchange of a homologous triplet from the initial binding state. This graph shows that the non-linearity makes adding additional triplets to the initial state is unfavorable once a mismatched triplet has bound, even when the subsequent base pairs are homologous.

\subsection{Possible Explanations of Biological Results}

We have already discussed the proposal that the energetic non-linearity explains why strand exchange is free energetically favorable even though the sequences of the incoming and outgoing strands are the same and the protein contacts in the initial searching ssDNA-RecA filament are similar to those in the final post-strand exchange state.

In addition, experimental results have shown that a rapid initial interaction incorporating  $\sim 15$  base pairs is followed by a slower progression of strand exchange that occurs in triplets \cite{Ragunathan}.  Figure~\ref{fig:nonlinearity} suggests that the binding of dsDNA to site II is favorable for fewer than 9 bases and requires only a few $kT$ of energy for fewer than 15 bases, whereas for more bases the binding is highly free energetically unfavorable.

Furthermore, FRET based studies indicate that homology recognition may be accurate for short sequences, but inaccurate for longer sequences \cite{Bazemore}.  A separate study showed that strand exchange pauses at sequence mismatches \cite{Lee,Sagi}, and we argue that such pauses lead to the unbinding of shorter non-homologous sequences because the binding of the dsDNA to the filament occurs sequentially  In the model presented here the pause in strand exchange at a mismatch results from the free energy cost of transferring the non-homologous triplet to the intermediate state as well as the cost of progressing past a mismatched triplet. Spontaneous unbinding of the entire strand exchange product becomes unlikely as the sequence lengthens because so many free energetically unfavorable transitions are required.   If the strand exchange product becomes too long, the unbinding time exceeds the recognition time available to the organism; however, as discussed above, accidental mismatches that extend beyond 18 bp 
rarely exist \textit{in vivo}. \textit{In vivo}, strand exchange does progress through regions of non-homology once a sufficiently long stand exchange product is formed, but ATP hydrolysis is required. \cite{Rosselli,Kim}

Finally, it is also well known that in the presence of ATP hydrolysis the size of the strand exchange product increases monotonically until it reaches a limit of $M \sim 80$ bp \cite{Dekker}, where $M$ is the number of bound dsDNA base pairs.  Strand exchange then continues to progress, but $M$ remains constant because the heteroduplex dsDNA unbinds from the lagging edge of the filament at the same average rate that new dsDNA binds to site II \cite{Dekker,Fan}.  Since the dsDNA can freely unbind from the filament, free energy minimization implies $M$ will remain $\sim M_{freemin}$. Additional effects associated with dynamics may explain why the strand exchange window moves along the dsDNA with $M \sim M_{freemin}$ rather than remaining stationary \cite{Conover}.  In contrast with the experimental results obtained in the presence of hydrolysis, experimental results obtained in the absence of hydrolysis show that the length of the strand exchange window can continue without bound.\cite{Dekker}  In the model 
presented here, if the number  of base pairs bound is small the mechanical energy penalty associated with adding base pairs is a quadratic function of energy for small numbers already bound; however, the base pairs redistribute the stress between the backbones, so that eventually the base pairs in the center of the filament are not under stress.  In this case, the penalty for adding another base pair triplet to the end becomes a linear function of the number bound rather than a quadratic function.  Thus, when a sufficient number of base pairs have undergone strand exchange, the energy required to add an additional triplet to the filament is constant, independent of the number bound.  If, as suggested by this model, the constant energy decrease due to additional DNA protein contacts is approximately equal to the constant energy increase associated with the added mechanical stress, then the filament can extend forever because both energies are independent of the number of base pairs already bound.

\subsection{Additional Features in Three Dimensions}

In the real RecA system steric factors are associated with the mismatch between the 150 bp persistence length of dsDNA and  the strong bending of dsDNA in the 18 bp/turn helical RecA filament.  The local rigidity of the dsDNA may play a role in limiting  the initial binding length to $\sim 9$ bp  since that many base pairs  can interact with the ssDNA-RecA without significant bending.
After some dsDNA triplets are bound, the rigidity may also play a role in preventing non-contiguous triplets from being added to the filament.  The nearest unbound triplet is already very near to the filament because it is attached to the bound triplets by the phosphate backbones which cannot extend much more than 0.5 nm/bp.   Thus, the phosphates are in a position to interact strongly with positively charged residues on the protein which can provide sufficient free energy for the required bending.    In contrast, the second neighboring triplet will be separated by a larger distance which reduces the interaction with the protein and requires more bending.  A detailed structural calculation would be required to correctly evaluate these effects, but both effects would further support rapid and accurate homology recognition.  In the simple one dimensional model discussed here, the free energy effects of the bending can be included in the $\gamma$ for the initial bound  state, but the additional degrees of 
freedom would alter the coupling between the initial bound state  $\gamma$ and the $\gamma$ for the intermediate state.

In addition, in the final post-strand exchange state interactions with the L1 and L2 loops may be more favorable for homologous triplets than non-homologous triplets due to steric factors.  Thus, the final state could have a sequence dependent linear contribution to the free energy that was not considered in this model, but may provide additional homology stringency.

\section{Conclusion}

We have proposed a simple mechanical model for the stress distribution on dsDNA bound to ssDNA-RecA filaments. The model suggests that a change in the conformation of one bound triplet can change the conformation of all of the other bound triplets; consequently, the total energy is a non-linear function of the number of bound base pairs.  The model  makes several significant qualitative and quantitative predictions.  The most important qualitative prediction is that a change in the configuration of one bound triplet changes configuration of all the other bound triplets.  The collective behavior of the triplets leads to a free energy that is a non-linear function of the number of consecutive bound triplets, where the binding of additional triplets becomes increasingly unfavorable as the number of bound triplets increases.  This non-linearity is important because in systems with more than $\sim 4 $ binding sites, neither thermodynamic equilibrium nor kinetic proofreading can combine accurate and  efficient homology recognition when the energy is a linear function of the number of correct pairings.  In contrast, an unfavorable non-linear energy combined with the a favorable linear energy due to DNA/protein contacts can promote rapid and accurate homology recognition by making initial sequence independent binding  interactions favorable for up to $\sim 9$ base pairs, while preventing any additional base pairs from binding unless the bound base pairs include 9 contiguous homologous base pairs.  If the initially bound base pairs do not contain 9 contiguous homologous base pairs, adding more  base pairs  to the filament is highly improbable regardless of whether or not the additional base pairs are homologous.   This effect combined with the statistics of the sequence distribution of bacterial genomes implies that of all but the $\sim$ 50 out of 10,000,000 possible pairings will rapidly unbind.

In addition , the non-linearity forces the addition of triplets to the filament to proceed sequentially from the initially binding, where adding more than two  base pair triplets after a mismatch is highly unlikely, even if the additional base pairs are sequence matched. Furthermore, the model suggests true homologs can proceed to complete strand exchange because the strand exchange of contiguous homologous base pair triplets reduces the tension on the dsDNA.  The tension reduction associated with the strand exchange of the initial $\sim 9$ contiguous homologous base pairs allows more base pairs to bind to the filament due to two effects: 1. the reduction in dsDNA tension reduces the free energy penalty for adding more triplets to the initial bound state 2. the decrease in the energy of the bound dsDNA due to strand exchange   reduces the unbinding rate of the dsDNA by making unbinding more unfavorable. As strand exchange progresses from the $\sim 9$ initial contiguous base pairs, the binding of the strand 
exchanged state still remains weak enough to be reversed unless $\sim 18$  contiguous bp make the transition to the final post-strand exchange state, a transition which is not favorable for $ <  \sim  18$ bp. These effects could provide exact sequence recognition for bacterial genomes except for repeated genes. The actual speed and accuracy of homology recognition  depends on detailed values of parameters as well as additional factors associated with the three dimensional geometry of the filament, so we do not provide detailed estimates here; however,  the simple model presented here provides mechanisms for overcoming fundamental limitations encountered in systems with more than 4 binding sites where the binding energy is a linear function of the number of correctly paired binding sites.

In sum, the energy non-linearity produces three crucial advantages that are unavailable to systems with a strictly linear energy: 1.  the initial interaction is limited to $\sim 15 $ bp beyond which binding of dsDNA triplets to the filament cannot progress without a sequence dependent transition of 9 contiguous homologous base pairs to the strand exchanged state 2.  nearly immediate unbinding of any sequence that does not contain at least 9 contiguous homologous base pairs  3. a large free energy penalty that prevents strand exchange from progressing past a sequence mismatch even if the mismatch is followed by homologous triplets 4.   a large free energy penalty that makes the transition from the intermediate state to the final state unfavorable until $\sim 18$ contiguous base pairs make the transition to the final state. These features provide much more rapid and accurate homology recognition than systems using linear energies: in systems with linear energies addition and strand exchange of a  homologous 
triplet is always favorable; therefore,  in systems with linear energies even short regions of accidental homology can produce substantial trapping times,  as demonstrated by both analytical modeling and numerical simulations.  \cite{Efraim} Qualitative features of the model provide possible explanations for well known but previously unexplained features for homology recognition and strand exchange and suggest that the bond rotations that appear in the overstretching of naked dsDNA may have a role in strand exchange.

\begin{acknowledgments}
We would like to thank Douglas Bishop, Yuen-Ling Chan, and Chantal Pr\'evost for helpful conversations about the X-ray structure of DNA bound to RecA.  We would also like to thank Chantal Pr\'evost for the PDB file of a complete RecA helix.
\end{acknowledgments}

\bibliography{deGennesbib}

\begin{thebibliography}{30}
\expandafter\ifx\csname natexlab\endcsname\relax\def\natexlab#1{#1}\fi
\expandafter\ifx\csname bibnamefont\endcsname\relax
  \def\bibnamefont#1{#1}\fi
\expandafter\ifx\csname bibfnamefont\endcsname\relax
  \def\bibfnamefont#1{#1}\fi
\expandafter\ifx\csname citenamefont\endcsname\relax
  \def\citenamefont#1{#1}\fi
\expandafter\ifx\csname url\endcsname\relax
  \def\url#1{\texttt{#1}}\fi
\expandafter\ifx\csname urlprefix\endcsname\relax\def\urlprefix{URL }\fi
\providecommand{\bibinfo}[2]{#2}
\providecommand{\eprint}[2][]{\url{#2}}

\bibitem[{\citenamefont{Roca and Cox}(1990)}]{Cox}
\bibinfo{author}{\bibfnamefont{A.}~\bibnamefont{Roca}} \bibnamefont{and}
  \bibinfo{author}{\bibfnamefont{M.}~\bibnamefont{Cox}}, \bibinfo{journal}{Mol.
  Biol.} \textbf{\bibinfo{volume}{25}}, \bibinfo{pages}{415}
  (\bibinfo{year}{1990}).

\bibitem[{\citenamefont{Kowalczykowski and Eggleston}(1994)}]{SteveK}
\bibinfo{author}{\bibfnamefont{S.}~\bibnamefont{Kowalczykowski}}
  \bibnamefont{and}
  \bibinfo{author}{\bibfnamefont{A.}~\bibnamefont{Eggleston}},
  \bibinfo{journal}{Annu. Rev. Biochem.} \textbf{\bibinfo{volume}{63}},
  \bibinfo{pages}{991} (\bibinfo{year}{1994}).

\bibitem[{\citenamefont{Chen et~al.}(2008)\citenamefont{Chen, Yang, and
  Pavletich}}]{Pavletich}
\bibinfo{author}{\bibfnamefont{Z.}~\bibnamefont{Chen}},
  \bibinfo{author}{\bibfnamefont{H.}~\bibnamefont{Yang}}, \bibnamefont{and}
  \bibinfo{author}{\bibfnamefont{N.}~\bibnamefont{Pavletich}},
  \bibinfo{journal}{Nature} \textbf{\bibinfo{volume}{453}},
  \bibinfo{pages}{489} (\bibinfo{year}{2008}).

\bibitem[{\citenamefont{Folta-Stogniew
  et~al.}(2004)\citenamefont{Folta-Stogniew, OÕMalley, Gupta, Anderson, and
  Radding}}]{Radding}
\bibinfo{author}{\bibfnamefont{E.}~\bibnamefont{Folta-Stogniew}},
  \bibinfo{author}{\bibfnamefont{S.}~\bibnamefont{OÕMalley}},
  \bibinfo{author}{\bibfnamefont{R.}~\bibnamefont{Gupta}},
  \bibinfo{author}{\bibfnamefont{K.}~\bibnamefont{Anderson}}, \bibnamefont{and}
  \bibinfo{author}{\bibfnamefont{C.}~\bibnamefont{Radding}},
  \bibinfo{journal}{Mol. Cell} \textbf{\bibinfo{volume}{15}},
  \bibinfo{pages}{965} (\bibinfo{year}{2004}).

\bibitem[{\citenamefont{Xiao et~al.}(2006)\citenamefont{Xiao, Lee, and
  Singleton}}]{Singleton}
\bibinfo{author}{\bibfnamefont{J.}~\bibnamefont{Xiao}},
  \bibinfo{author}{\bibfnamefont{A.}~\bibnamefont{Lee}}, \bibnamefont{and}
  \bibinfo{author}{\bibfnamefont{S.}~\bibnamefont{Singleton}},
  \bibinfo{journal}{ChemBioChem} \textbf{\bibinfo{volume}{7}},
  \bibinfo{pages}{1265} (\bibinfo{year}{2006}).

\bibitem[{\citenamefont{Mazin and Kowalczykowski}(1996)}]{Mazin}
\bibinfo{author}{\bibfnamefont{A.}~\bibnamefont{Mazin}} \bibnamefont{and}
  \bibinfo{author}{\bibfnamefont{S.}~\bibnamefont{Kowalczykowski}},
  \bibinfo{journal}{Proc. Natl. Acad. Sci. USA} \textbf{\bibinfo{volume}{93}},
  \bibinfo{pages}{10673} (\bibinfo{year}{1996}).

\bibitem[{\citenamefont{Menetski et~al.}(1990)\citenamefont{Menetski, Bear, and
  Kowalczykowski}}]{Menetski}
\bibinfo{author}{\bibfnamefont{J.}~\bibnamefont{Menetski}},
  \bibinfo{author}{\bibfnamefont{D.}~\bibnamefont{Bear}}, \bibnamefont{and}
  \bibinfo{author}{\bibfnamefont{S.}~\bibnamefont{Kowalczykowski}},
  \bibinfo{journal}{Proc. Natl. Acad. Sci. USA} \textbf{\bibinfo{volume}{87}},
  \bibinfo{pages}{21} (\bibinfo{year}{1990}).

\bibitem[{\citenamefont{Rosselli and Stasiak}(1990)}]{Stasiak}
\bibinfo{author}{\bibfnamefont{W.}~\bibnamefont{Rosselli}} \bibnamefont{and}
  \bibinfo{author}{\bibfnamefont{A.}~\bibnamefont{Stasiak}},
  \bibinfo{journal}{J. Mol. Biol.} \textbf{\bibinfo{volume}{216}},
  \bibinfo{pages}{335} (\bibinfo{year}{1990}).

\bibitem[{\citenamefont{Stasiak et~al.}(1981)\citenamefont{Stasiak, Capua, and
  Koller}}]{Koller}
\bibinfo{author}{\bibfnamefont{A.}~\bibnamefont{Stasiak}},
  \bibinfo{author}{\bibfnamefont{E.~D.} \bibnamefont{Capua}}, \bibnamefont{and}
  \bibinfo{author}{\bibfnamefont{T.}~\bibnamefont{Koller}},
  \bibinfo{journal}{J. Mol. Biol.} \textbf{\bibinfo{volume}{151}},
  \bibinfo{pages}{557} (\bibinfo{year}{1981}).

\bibitem[{\citenamefont{Savir and Tlusty}(2010)}]{Tlusty}
\bibinfo{author}{\bibfnamefont{Y.}~\bibnamefont{Savir}} \bibnamefont{and}
  \bibinfo{author}{\bibfnamefont{T.}~\bibnamefont{Tlusty}},
  \bibinfo{journal}{Mol. Cell} \textbf{\bibinfo{volume}{40}},
  \bibinfo{pages}{388} (\bibinfo{year}{2010}).

\bibitem[{\citenamefont{Mazin and Kowalczykowski}(1999)}]{MazinSteveK}
\bibinfo{author}{\bibfnamefont{A.}~\bibnamefont{Mazin}} \bibnamefont{and}
  \bibinfo{author}{\bibfnamefont{S.}~\bibnamefont{Kowalczykowski}},
  \bibinfo{journal}{Genes Dev.} \textbf{\bibinfo{volume}{13}},
  \bibinfo{pages}{2005} (\bibinfo{year}{1999}).

\bibitem[{\citenamefont{Klapstein et~al.}(2004)\citenamefont{Klapstein, Chou,
  and Bruinsma}}]{Bruinsma}
\bibinfo{author}{\bibfnamefont{K.}~\bibnamefont{Klapstein}},
  \bibinfo{author}{\bibfnamefont{T.}~\bibnamefont{Chou}}, \bibnamefont{and}
  \bibinfo{author}{\bibfnamefont{R.}~\bibnamefont{Bruinsma}},
  \bibinfo{journal}{Biophys J.} \textbf{\bibinfo{volume}{87}},
  \bibinfo{pages}{1466} (\bibinfo{year}{2004}).

\bibitem[{\citenamefont{Kates-Harbeck et~al.}()\citenamefont{Kates-Harbeck,
  Tilloy, and Prentiss}}]{Julian}
\bibinfo{author}{\bibfnamefont{J.}~\bibnamefont{Kates-Harbeck}},
  \bibinfo{author}{\bibfnamefont{A.}~\bibnamefont{Tilloy}}, \bibnamefont{and}
  \bibinfo{author}{\bibfnamefont{M.}~\bibnamefont{Prentiss}}, \bibinfo{note}{to
  be published}.

\bibitem[{\citenamefont{Feinstein and Prentiss}()}]{Efraim}
\bibinfo{author}{\bibfnamefont{E.}~\bibnamefont{Feinstein}} \bibnamefont{and}
  \bibinfo{author}{\bibfnamefont{M.}~\bibnamefont{Prentiss}}, \bibinfo{note}{to
  be published}.

\bibitem[{\citenamefont{Hsieh et~al.}(1992)\citenamefont{Hsieh, Camerini-Otero,
  and Camerini-Otero}}]{Hsieh}
\bibinfo{author}{\bibfnamefont{P.}~\bibnamefont{Hsieh}},
  \bibinfo{author}{\bibfnamefont{C.}~\bibnamefont{Camerini-Otero}},
  \bibnamefont{and}
  \bibinfo{author}{\bibfnamefont{R.}~\bibnamefont{Camerini-Otero}},
  \bibinfo{journal}{Proc. Natl Acad. Sci. USA} \textbf{\bibinfo{volume}{89}},
  \bibinfo{pages}{6492} (\bibinfo{year}{1992}).

\bibitem[{\citenamefont{Hopfield}(1974)}]{Hopfield}
\bibinfo{author}{\bibfnamefont{J.}~\bibnamefont{Hopfield}},
  \bibinfo{journal}{Proc. Natl. Acad. Sci. USA} \textbf{\bibinfo{volume}{71}},
  \bibinfo{pages}{4135} (\bibinfo{year}{1974}).

\bibitem[{\citenamefont{Bazemore et~al.}(1997)\citenamefont{Bazemore,
  Folta-Stogniew, Takahashi, and Radding}}]{Bazemore}
\bibinfo{author}{\bibfnamefont{L.}~\bibnamefont{Bazemore}},
  \bibinfo{author}{\bibfnamefont{E.}~\bibnamefont{Folta-Stogniew}},
  \bibinfo{author}{\bibfnamefont{M.}~\bibnamefont{Takahashi}},
  \bibnamefont{and} \bibinfo{author}{\bibfnamefont{C.}~\bibnamefont{Radding}},
  \bibinfo{journal}{Proc. Natl. Acad. Sci. USA} \textbf{\bibinfo{volume}{94}},
  \bibinfo{pages}{11863} (\bibinfo{year}{1997}).

\bibitem[{\citenamefont{Peacock-Villada
  et~al.}(2012)\citenamefont{Peacock-Villada, Yang, Danilowicz, Feinstein,
  Pollack, McShan, Coljee, and Prentiss}}]{Alex}
\bibinfo{author}{\bibfnamefont{A.}~\bibnamefont{Peacock-Villada}},
  \bibinfo{author}{\bibfnamefont{D.}~\bibnamefont{Yang}},
  \bibinfo{author}{\bibfnamefont{C.}~\bibnamefont{Danilowicz}},
  \bibinfo{author}{\bibfnamefont{E.}~\bibnamefont{Feinstein}},
  \bibinfo{author}{\bibfnamefont{N.}~\bibnamefont{Pollack}},
  \bibinfo{author}{\bibfnamefont{S.}~\bibnamefont{McShan}},
  \bibinfo{author}{\bibfnamefont{V.}~\bibnamefont{Coljee}}, \bibnamefont{and}
  \bibinfo{author}{\bibfnamefont{M.}~\bibnamefont{Prentiss}},
  \bibinfo{journal}{Nucl. Acids Res.} \textbf{\bibinfo{volume}{40}},
  \bibinfo{pages}{10441} (\bibinfo{year}{2012}).

\bibitem[{\citenamefont{Ragunathan et~al.}(2011)\citenamefont{Ragunathan, Joo,
  and Ha}}]{Ragunathan}
\bibinfo{author}{\bibfnamefont{K.}~\bibnamefont{Ragunathan}},
  \bibinfo{author}{\bibfnamefont{C.}~\bibnamefont{Joo}}, \bibnamefont{and}
  \bibinfo{author}{\bibfnamefont{T.}~\bibnamefont{Ha}},
  \bibinfo{journal}{Structure} \textbf{\bibinfo{volume}{19}},
  \bibinfo{pages}{1064} (\bibinfo{year}{2011}).

\bibitem[{\citenamefont{Karplus}()}]{Karplus}
\bibinfo{author}{\bibfnamefont{M.}~\bibnamefont{Karplus}},
  \bibinfo{note}{private communication}.

\bibitem[{\citenamefont{Danilowicz et~al.}(2011)\citenamefont{Danilowicz,
  Feinstein, Conover, Coljee, Vlassakis, Chan, Bishop, and
  Prentiss}}]{Danilowicz}
\bibinfo{author}{\bibfnamefont{C.}~\bibnamefont{Danilowicz}},
  \bibinfo{author}{\bibfnamefont{E.}~\bibnamefont{Feinstein}},
  \bibinfo{author}{\bibfnamefont{A.}~\bibnamefont{Conover}},
  \bibinfo{author}{\bibfnamefont{V.}~\bibnamefont{Coljee}},
  \bibinfo{author}{\bibfnamefont{J.}~\bibnamefont{Vlassakis}},
  \bibinfo{author}{\bibfnamefont{Y.}~\bibnamefont{Chan}},
  \bibinfo{author}{\bibfnamefont{D.}~\bibnamefont{Bishop}}, \bibnamefont{and}
  \bibinfo{author}{\bibfnamefont{M.}~\bibnamefont{Prentiss}},
  \bibinfo{journal}{Nucl. Acids Res.} \textbf{\bibinfo{volume}{40}},
  \bibinfo{pages}{1717} (\bibinfo{year}{2011}).

\bibitem[{\citenamefont{van~der Heijden et~al.}(2008)\citenamefont{van~der
  Heijden, Modesti, Hage, Kanaar, Wyman, and Dekker}}]{Dekker}
\bibinfo{author}{\bibfnamefont{T.}~\bibnamefont{van~der Heijden}},
  \bibinfo{author}{\bibfnamefont{M.}~\bibnamefont{Modesti}},
  \bibinfo{author}{\bibfnamefont{S.}~\bibnamefont{Hage}},
  \bibinfo{author}{\bibfnamefont{R.}~\bibnamefont{Kanaar}},
  \bibinfo{author}{\bibfnamefont{C.}~\bibnamefont{Wyman}}, \bibnamefont{and}
  \bibinfo{author}{\bibfnamefont{C.}~\bibnamefont{Dekker}},
  \bibinfo{journal}{Mol. Cell} \textbf{\bibinfo{volume}{30}},
  \bibinfo{pages}{530} (\bibinfo{year}{2008}).

\bibitem[{\citenamefont{Gennes}(2001)}]{deGennes}
\bibinfo{author}{\bibfnamefont{P.~G.~D.} \bibnamefont{Gennes}},
  \bibinfo{journal}{CR. Acad. Sci. IV-Phys.} \textbf{\bibinfo{volume}{2}},
  \bibinfo{pages}{1505} (\bibinfo{year}{2001}).

\bibitem[{\citenamefont{{SantaLucia, Jr.}}(1998)}]{SantaLucia}
\bibinfo{author}{\bibfnamefont{J.}~\bibnamefont{{SantaLucia, Jr.}}},
  \bibinfo{journal}{Proc. Natl Acad. Sci. USA} \textbf{\bibinfo{volume}{95}},
  \bibinfo{pages}{1460} (\bibinfo{year}{1998}).

\bibitem[{\citenamefont{Lee et~al.}(2006)\citenamefont{Lee, Xiao, and
  Singleton}}]{Lee}
\bibinfo{author}{\bibfnamefont{A.}~\bibnamefont{Lee}},
  \bibinfo{author}{\bibfnamefont{J.}~\bibnamefont{Xiao}}, \bibnamefont{and}
  \bibinfo{author}{\bibfnamefont{S.}~\bibnamefont{Singleton}},
  \bibinfo{journal}{J. Mol. Biol.} \textbf{\bibinfo{volume}{360}},
  \bibinfo{pages}{343} (\bibinfo{year}{2006}).

\bibitem[{\citenamefont{D.~Sagi~and? T.~Tlusty}(2006)}]{Sagi}
\bibinfo{author}{\bibfnamefont{a.~S.} \bibnamefont{D.~Sagi~and? T.~Tlusty}},
  \bibinfo{journal}{Nucl. Acids Res.} \textbf{\bibinfo{volume}{34}},
  \bibinfo{pages}{5021} (\bibinfo{year}{2006}).

\bibitem[{\citenamefont{Rosselli and Stasiak}(1991)}]{Rosselli}
\bibinfo{author}{\bibfnamefont{W.}~\bibnamefont{Rosselli}} \bibnamefont{and}
  \bibinfo{author}{\bibfnamefont{A.}~\bibnamefont{Stasiak}},
  \bibinfo{journal}{EMBO J.} \textbf{\bibinfo{volume}{10}},
  \bibinfo{pages}{4391} (\bibinfo{year}{1991}).

\bibitem[{\citenamefont{Kim et~al.}(1992)\citenamefont{Kim, Cox, and
  Inman}}]{Kim}
\bibinfo{author}{\bibfnamefont{J.}~\bibnamefont{Kim}},
  \bibinfo{author}{\bibfnamefont{M.}~\bibnamefont{Cox}}, \bibnamefont{and}
  \bibinfo{author}{\bibfnamefont{R.}~\bibnamefont{Inman}}, \bibinfo{journal}{J.
  Biol. Chem.} \textbf{\bibinfo{volume}{267}}, \bibinfo{pages}{16444}
  (\bibinfo{year}{1992}).

\bibitem[{\citenamefont{Fan et~al.}(2011)\citenamefont{Fan, Cox, and Li}}]{Fan}
\bibinfo{author}{\bibfnamefont{H.-F.} \bibnamefont{Fan}},
  \bibinfo{author}{\bibfnamefont{M.}~\bibnamefont{Cox}}, \bibnamefont{and}
  \bibinfo{author}{\bibfnamefont{H.-W.} \bibnamefont{Li}},
  \bibinfo{journal}{PLoS ONE} \textbf{\bibinfo{volume}{6}},
  \bibinfo{pages}{e21359} (\bibinfo{year}{2011}).

\bibitem[{\citenamefont{Conover et~al.}(2011)\citenamefont{Conover, Danilowicz,
  Gunaratne, Coljee, Kleckner, and Prentiss}}]{Conover}
\bibinfo{author}{\bibfnamefont{A.}~\bibnamefont{Conover}},
  \bibinfo{author}{\bibfnamefont{C.}~\bibnamefont{Danilowicz}},
  \bibinfo{author}{\bibfnamefont{R.}~\bibnamefont{Gunaratne}},
  \bibinfo{author}{\bibfnamefont{V.}~\bibnamefont{Coljee}},
  \bibinfo{author}{\bibfnamefont{N.}~\bibnamefont{Kleckner}}, \bibnamefont{and}
  \bibinfo{author}{\bibfnamefont{M.}~\bibnamefont{Prentiss}},
  \bibinfo{journal}{Nucl. Acids Res.} \textbf{\bibinfo{volume}{39}},
  \bibinfo{pages}{8833} (\bibinfo{year}{2011}).

\end{thebibliography}

\end{document}